# Differential rotation of the solar transition region from STEREO/EUVI 30.4 nm images

Jaidev Sharma[1*], Brajesh Kumar [2], Anil K Malik [1] and Hari Om Vats[3]

[1]*Department of Physics, C.C.S. University, Meerut, 201001, India*

[2]*Udaipur Solar Observatory, Physical Research Laboratory, Dewali, Badi Road, Udaipur 313004 Rajasthan, India.*

[3]*Space Education and Research Foundation, Ahmedabad 380054, India*

**ABSTRACT**

The solar photosphere, chromosphere and corona are known to rotate differentially as a function of latitude. To date, it is unclear if the solar transition region also rotates differentially. In this paper, we investigate differential rotational profile of solar transition region as a function of latitude, using solar full disk (SFD) images at 30.4 nm wavelength recorded by Extreme Ultraviolet Imager (EUVI) of onboard *Solar Terrestrial Relations Observatory (STEREO)* space mission for the period from 2008 to 2018 (Solar Cycle 24). Our investigations show that solar transition region rotates differentially. The sidereal rotation rate obtained at $\pm 5$ deg. equatorial band is quite high (~ 14.7 deg./day), which drops to ~ 13.6 deg./day towards both polar regions. We also obtain that the rotational differentiality is low during the period of high solar activity (rotation rate varies from 14.86 to 14.27 deg./day) while it increases during the ascending and the descending phases of the 24$^{th}$ solar cycle (rotation rate varies from 14.56 to ~ 13.56 deg./day in 2008 and 14.6 to 13.1 deg./day in 2018). Average sidereal rotation rate (over SFD) follows the trend of solar activity (maximum ~ 14.97 deg./day during the peak phase of the solar activity, which slowly decreases to minimum ~ 13.9 deg./day during ascending and the descending phases of the 24$^{th}$ solar cycle). We also observe that solar transition region rotates less differentially than the corona.

## 1 INTRODUCTION

The study of the differential profile of solar rotation is crucial to understand the behavior of solar dynamo process. It is noteworthy to mention that the rotation of solar interior with respect to latitude as well as altitude plays important role in the process of the twisting of solar poloidal magnetic fields, which results into the birth of sunspots on the solar photosphere. Techniques to obtain rotational profile of solar interior are well developed using the tools of helioseismology (Brown **1985**; Thompson *et. al.* **2003**; Howe **2009**). However, there are three main methods to obtain the differential rotation of solar atmosphere, and these are as follows: feature tracking approach (Gilman **1974**), spectroscopic measurements (Howard **1984**) and flux modulation method (Vats *et. al.* **2001**). All aforementioned techniques have their own constraints and errors. Solar astronomers are continuously working to improve these techniques by minimizing estimation errors for yielding better results and thereby precious measurement of solar rotation.

Sime *et. al.* (**1989**) used the synoptic observations of the coronal Fe $_{XIV}$ lines recorded by Sacramento Peak 40 cm coronagraph (for the period from 1973 to 1985) corresponding to 1.15 R$_\odot$ at 5303 Å to study the synodic rotation period of latitudinal bands. They reported that the synodic rotation period of latitudinal band at $\pm$ 30 deg. of the Fe $_{XIV}$ green line corona is 27.5 days. Moreover, Fe $_{XIV}$ green line corona rotates less differentially as compared to photospheric and chromospheric features. Several researchers (Newton & Nunn **1951**; Howard, Gilman & Gilman **1984**; Balthasar, Vazquez & Woehl **1986**; Howard **1991**, **1996** and Shivraman *et. al.* **1993**) investigated solar rotation phenomena using tracer-tracking method. Their results show that various features of the Sun viz., sunspots (SSNs), plages, faculae, coronal bright points (CBPs), giant cells and filaments etc. rotate with the Sun itself. Timothy, Krieger & Vaiana **1975**; Weber & Sturrock **2002**; Kariyappa **2008**; and Chandra, Vats & Iyer **2009** reported mixed behavior (either rigid or differential nature) of the latitudinal rotational profile of the solar corona as seen in the soft X-ray observations. These results clearly indicate that the actual picture of the rotational profile of solar corona is not clear yet. Weber & Sturrock (**2002**) also reported rigid rotational behavior of corona using images observed from soft X-ray telescope (SXT) onboard *Yohkoh* than that of the lower atmosphere of the Sun. Kariyappa (**2008**) used observations from the *Yohkoh* and *Hinode* solar space missions and also reported that the corona exhibits phase independent nature of differential rotation. Chandra *et. al.* (**2009**) used flux modulation technique on solar full disk (SFD) images recorded by the Nobeyama Radioheliograph (NoRH) telescope at 17 GHz for the period from 1990 to 2001.They reported that the rotation periods of solar corona are in phase with solar activity, whereas gradient of rotation exists in anti-phase with respect to the solar activity indicators. Sharma *et. al.* (**2020a**) used the *Solar Dynamics Observatory* (SDO) observations at 9.4 nm, 13.1 nm, 17.1 nm, 19.3 nm, 21.1 nm and 33.5 nm for the period from 2012 to 2018. They reported yearly trend, band wise trend [(- 60, - 50, - 40), (- 10, 0, 10) and (40, 50, 60)] and average latitudinal trend (average from - 60 to + 60) in rotation periods decrease with respect to the increasing temperature of the solar coronal layers.

The solar transition region is a thin and very irregular layer of solar atmosphere that lies between chromosphere and corona having the temperature ranging from 20,000 K to 1,000,000 K (Mariska, **1993**). Gallagher *et. al.* (1998) used the observations of the quiet Sun network recorded by Coronal Diagnostic Spectrometer (CDS) onboard *Solar and Heliospheric Observatory* (*SOHO)* space mission. They found that the concentrations of network areas and emission of solar transition region are high as compare to the chromosphere and corona. This layer is difficult to observe by ground-based observatories because it contains the ionized atoms like hydrogen, oxygen, carbon, and silicon that emit radiation in extreme ultraviolet (EUV) range. Some space based observatories like *Solar Maximum Mission (SMM)*, *Solar and Heliospheric Observatory (SOHO)*, *Solar Terrestrial Relations Observatory (STEREO)* and *Transition Region and Coronal Explorer (TRACE)* have been extracting information regarding the structure and dynamics of the transition region. To the best of our knowledge, the rotational profile of the transition region could not be investigated because it is thin and irregular in nature. The irregularities in the transition region may have introduced some severe limitations in the previously applied techniques.

In this paper, we investigate the rotational profile of the solar transition region using flux modulation method. The method is applied to the *STEREO-A* observations at 30.4 nm for the period from 2008 to 2018 (i.e., Solar Cycle 24). We made sixteen continuous bands on SFD (Fig. **1**) to extract flux or intensity, which is auto correlated up to a lag of 150 days to check the similarity in rotational signals. The first secondary maxima of autocorrelograms are fitted by Gaussian function to filter out the noise. The time value (along horizontal axis) of the Gaussian peak gives the synodic rotation period in days. We investigate the differential rotational profile of the transition region for all latitudinal bands. We also investigate the correlation between average rotation rates of SFD images in the solar transition region with the sunspot numbers. Our investigations show that the rotation rate in the equatorial region is quite high, which drops towards the poles. Furthermore, average rotation rate (deg./day) and annual sunspot numbers are correlated in the same phase. Hence, we conclude that the average rotation rate of the solar transition region follows the solar activity during the Solar Cycle 24. In Section 2, the observations and methods are described, while the results and related discussions are presented in Section 3. Finally, in Section 4 we present the conclusions arrived at in this work.

**2 OBSERVATIONS AND METHODOLOGY**

The full disk images of the solar transition region corresponding to 30.4 nm are obtained from Extreme Ultraviolet Imager (EUVI) instrument onboard the *Solar Terrestrial Relations Observatory (STEREO-A)* space mission (NASA mission operational since October 2006). The *STEREO* observatory comprises of two spacecrafts, *STEREO-A* and *STEREO-B* aimed to obtain the full-disk images of the Sun at the wavelengths 30.4 nm, 17.1 nm, 19.5 nm and 28.4 nm (Wuelser *et. al.* 2004). The ground connection with *STEREO-B* has been lost in 2014 while *STEREO-A* is still observing the Sun. We used observations recorded by *STEREO-A* instrument at 30.4 nm for the duration from 2008 to 2018 to investigate the differential rotational behavior of the solar transition region. During our investigations, we exclude data obtained in the year 2009 due to poor features, which results in insufficient rotational information. We also exclude the data for almost first seven months in the year 2015 due to unavailability of the data (during this period, there were problems in the spacecraft communication system). However, the interpolated data are used for the year 2009 and 2015 as shown in the Figs. **4**, **5** & **6**. We selected continuous rectangular latitudinal bands containing entire pixels on SFD along length (Fig.**1**).The width of the bands decreases with increasing latitudes on both the hemispheres.

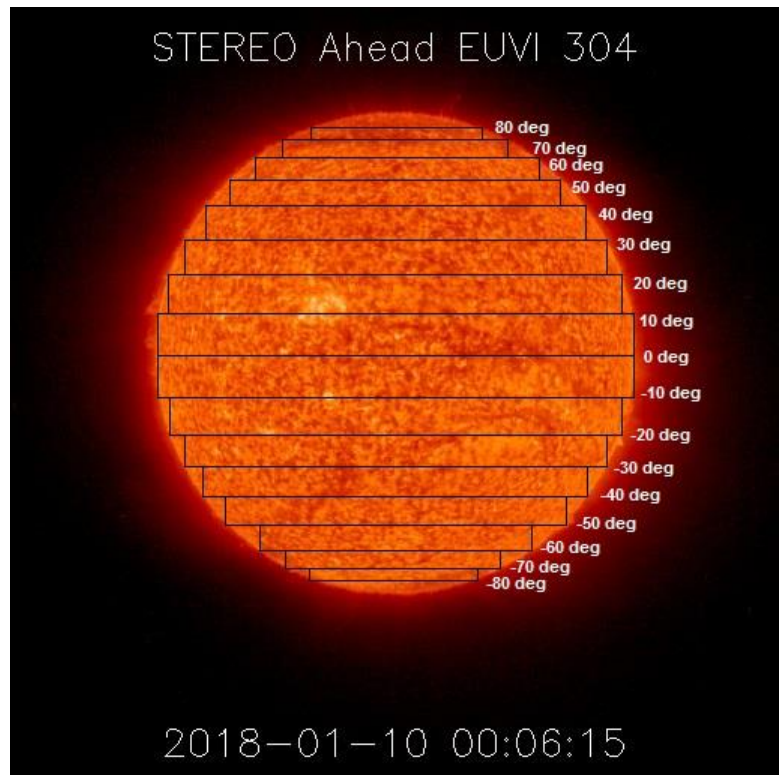

**Figure 1:** Schematic of rectangular latitudinal bands selected at equal interval of 10 deg. on SFD image at 30.4 nm obtained from *STEREO-A* space mission.

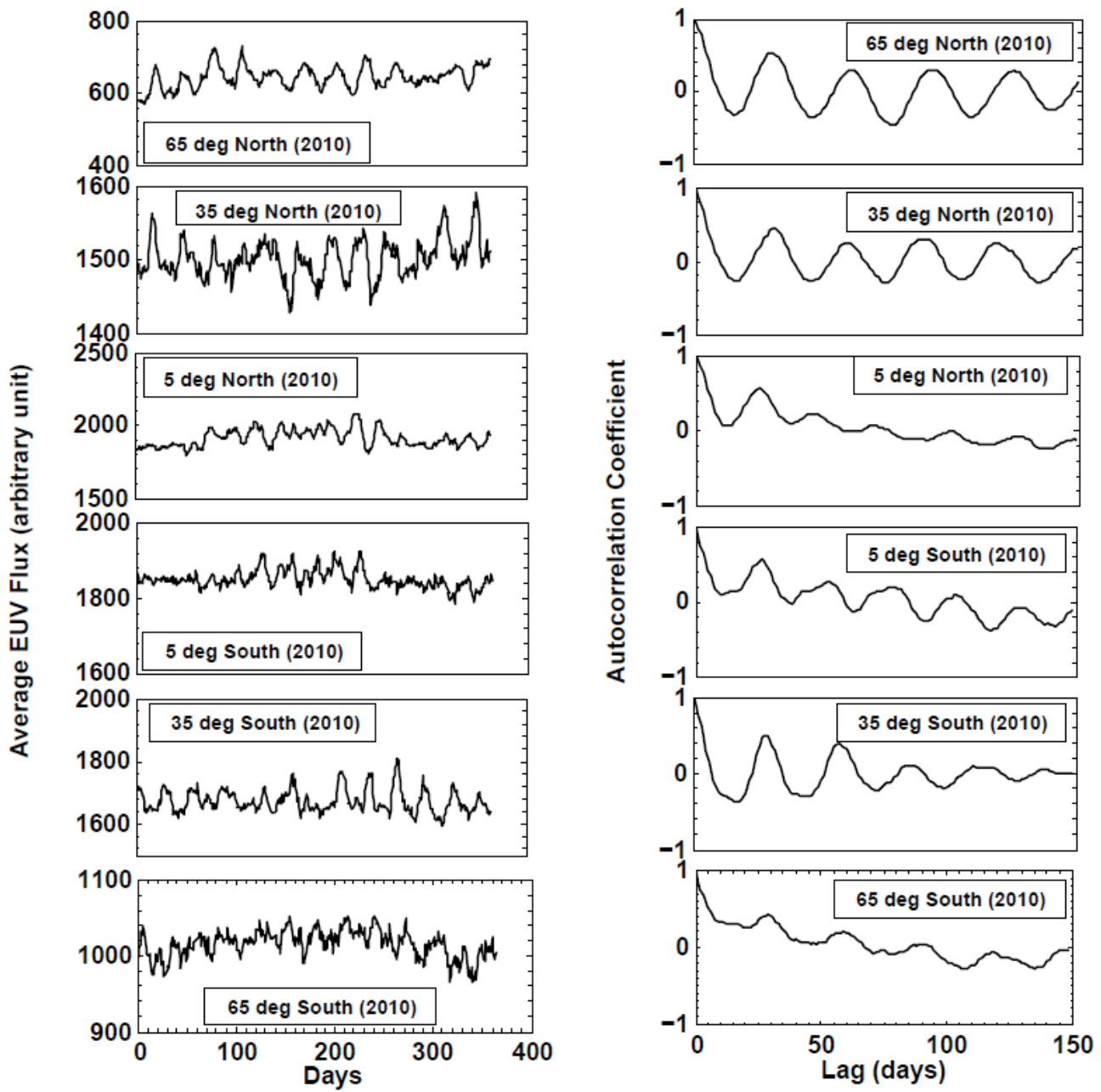

**Figure 2:** The left panels show plots of average EUV flux/intensity as a function of time for equatorial, mid and high latitudinal bands for both sides of the equator using 30.4 nm observations from *STEREO-A* in the year 2010, whereas the right panels show their corresponding autocorrelograms, which contain the periodicities in rotational features.

Sixteen latitudinal bands with widths of 10 deg. were extracted from - 80 to + 80 deg. as shown in Fig. 1. Insufficient features are observed over the bands (- 80 to - 70) and (70 to 80), therefore, we do not consider these bands in our investigations. The average flux is extracted from latitudinal bands at - 65, - 55, - 45, - 35, - 25, - 15, - 5, 5, 15, 25, 35, 45, 55, and 65 deg. on both the hemispheres (negative and positive signs indicate latitudes on southern and northern hemisphere, respectively). These variable bands cover more pixels on equatorial region (hence, comprises of sufficient features) and successively less pixels corresponding to increasing latitudinal bands on both the hemispheres (hence, comprises of less content of noise). Nevertheless, we found significant signals of rotational features with tolerable amount of noise, which results in the statistically significant peaks of autocorrelograms. In Fig. 2, the left panels show temporal variation of average EUV flux/intensity for equatorial, mid and high latitudinal bands on both sides of the hemispheres in the year 2010 whereas the right panels show their corresponding autocorrelograms (up to a lag of 150 days) that contain the evidence of periodicities in rotational features. The first secondary maxima of autocorrelograms are much higher than the rest, which shows more statistical significance in comparison to rest of the maxima owing to the higher value of correlation coefficient (c.f., Fig. 2). Hence, in order to obtain accurate rotational properties, the first secondary maximum is used. It is reasonable to fit the Gaussian function at the first secondary maximum of autocorrelogram (Sharma *et. al.* **2020a**) in order to filter the noise content. Thus, Gaussian function has been fitted on first secondary maximum of autocorrelogram to enhance the symmetric behavior, which finally results into the better accuracy in the measurements of rotation periods (c.f., Fig **3**). The time value (in days) of the peak of the symmetric Gaussian function (along horizontal axis) gives the synodic rotation period in days.

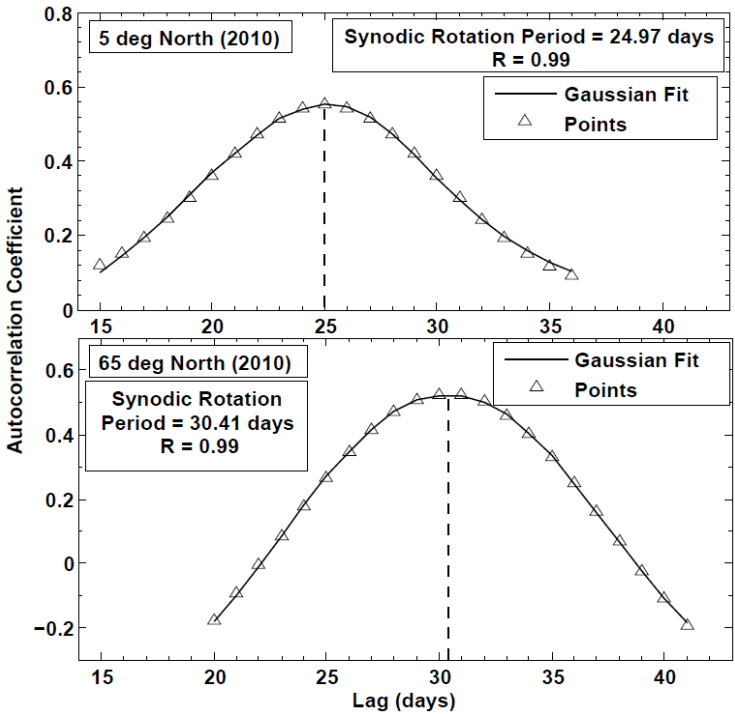

**Figure 3:** Peaks of the first secondary maxima of the autocorrelograms (triangles) corresponding to 5 deg. North (Top panel) and 65 deg. North (Bottom panel) in the year 2010 using *STEREO-A* (30.4 nm) observations fitted with Gaussian function as shown by continuous line. Vertical dashed line corresponds to the central value of Gaussian function.

In Fig.3, Gaussian functions (continuous line) are fitted on first secondary maximum (triangles) of autocorrelograms at 5 deg. North (top panel) and 65 deg. North (bottom panel) for the year 2010, which shows a considerable variation in rotation periods for lower and higher latitudes. The vertical dashed lines are plotted corresponding to the central value (synodic rotation period in days) of Gaussian function. The Pearson's coefficient of Gaussian fitting is obtained as 0.99, which shows high level of accuracy in the estimation of synodic rotation periods. Sidereal rotation period is obtained as

$$T_{sidereal} = \frac{346\, T_{synodic}}{346 + T_{synodic}} \quad (1)$$

Here, orbital period of STEREO-A instrument is 346 days.

The relation between angular velocity $\omega_{rot}$ (in deg./day) and rotation rate T (in days) is obtained as

$$\omega_{rot} = \frac{360°}{T} \quad (2)$$

Several researchers (Vats and Chandra, **2011**, Sharma *et. al.* **2020b**) have reported North-South asymmetry in solar rotation. So, we used asymmetric equation to calculate the rotational parameters as

$$\omega_{rot}(\phi) = C_0 + C_1\, Sin\phi + C_2\, Sin^2\phi + C_3\, Sin^3\phi + C_4\, Sin^4\phi \quad (3)$$

where $\phi$ is the latitude in degree.

**Table 1:** Coefficients (deg./day) of the solar rotational profile with their standard errors (SE). The last column shows the root mean square error (RMSE) in the fitting process.

| Year | $C_0 \pm SE$ | $C_1 \pm SE$ | $C_2 \pm SE$ | RMSE |
|---|---|---|---|---|
| 2008 | 14.6 ± 0.3 | -0.1 ± 0.3 | -1.09 ± 0.6 | 0.5 |
| 2010 | 14.5 ± 0.4 | -0.7 ± 0.4 | -1.9 ± 0.8 | 0.6 |
| 2011 | 14.9 ± 0.4 | -0.3 ± 0.4 | -2.7 ± 0.8 | 0.8 |
| 2012 | 14.9 ± 0.1 | 0.7 ± 0.6 | -1.4 ± 0.1 | 0.2 |
| 2013 | 14.4 ± 0.2 | -0.2 ± 0.7 | -0.77 ± 0.7 | 0.4 |
| 2014 | 14.8 ± 0.2 | 0.03 ± 0.2 | -0.7 ± 0.5 | 0.4 |
| 2016 | 14.6 ± 0.08 | 0.05 ± 0.1 | -0.6 ± 0.2 | 0.1 |
| 2017 | 14.6 ± 0.2 | 0.04 ± 0.2 | -0.86 ± 0.5 | 0.3 |
| 2018 | 14.6 ± 0.1 | 0.47 ± 0.2 | -1.4 ± 0.4 | 0.2 |
| **Avg** | 14.7 ± 0.26 | 0.001 ± 0.3 | -1.26 ± 0.5 | 0.4 |

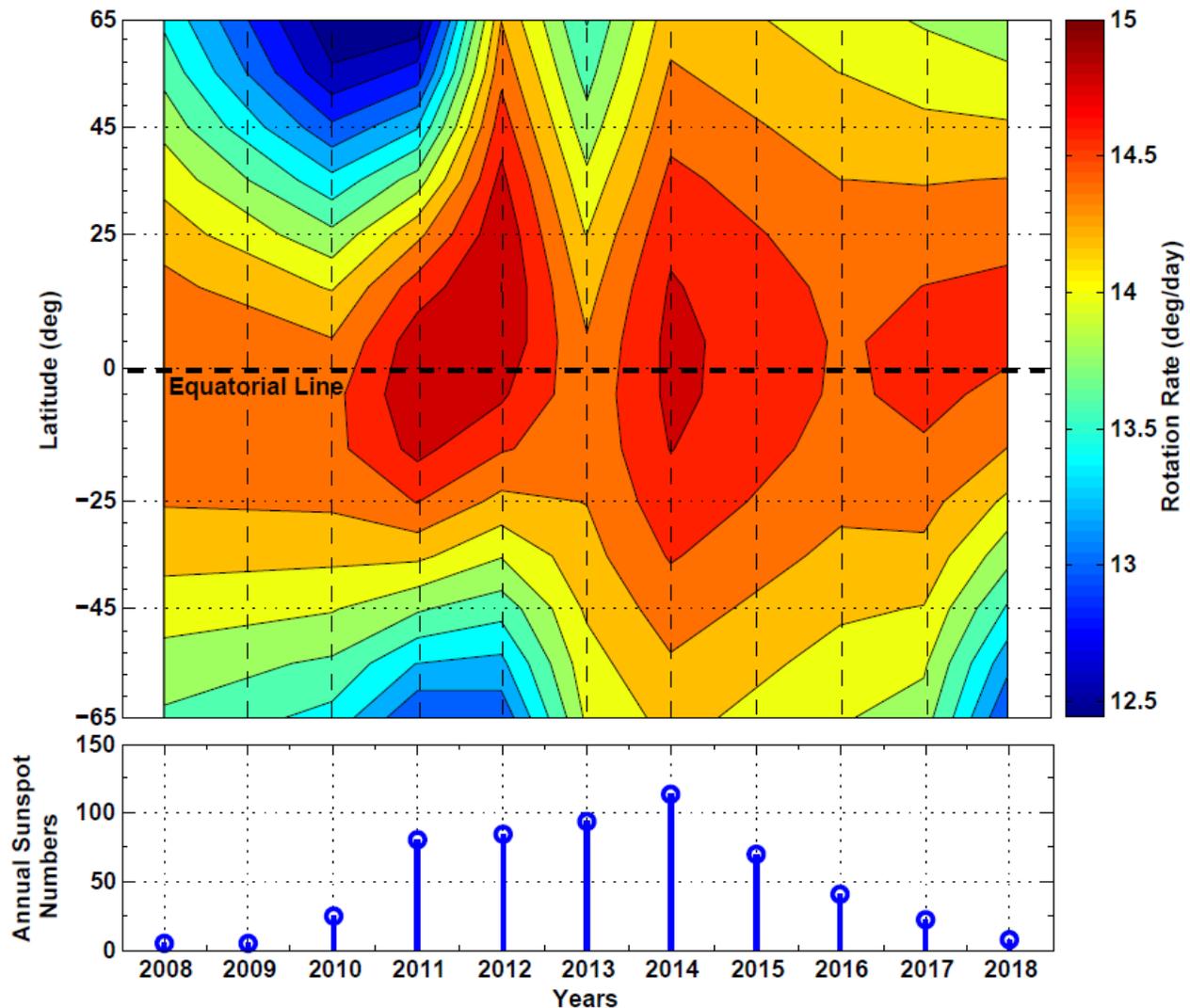

**Figure 4:** Contour plot (top panel) shows temporal (along horizontal axis) and latitudinal (along vertical axis) variation of the rotation rate (shown in color code) of the solar transition region in deg./day. The bottom panel shows the temporal variation of annual sunspot numbers for the period from 2008 to 2018.

In Eq. (3), $C_0$ (deg./day) represents the equatorial rotation rate, $C_1$ (deg./day) and $C_3$ (deg./day) are respective asymmetry indicators at middle and higher latitudes whereas $C_2$ (deg./day) and $C_4$ (deg./day) corresponds to the gradients at the middle and the upper latitudes, respectively. In order to remove the crosstalk, we substitute $C_4 = 0$ in Eq. (3) (Sharma *et. al.* **2020a**). Since this study is limited up to the middle latitudes (± 65 deg.) on both side of the hemispheres, therefore, it is reasonable to substitute $C_3 = 0$ in Eq. (3). The calculated coefficients $C_0$, $C_1$, and $C_2$ are given in the table 1 for the period of 2008- 2018 (except for the year 2009 and 2015). The last column contains root mean square error (RMSE) of the fitting.

**3 RESULTS AND DISCUSSIONS**

The rotation rate (Eq. 2) is shown in the contour plot (Fig. 4), which depicts the yearly latitudinal profile of sidereal rotation rate (in deg./day). Fig.4 shows that time average of sidereal rotation rate (from 2008 to 2018) of the equatorial region (- 5 to + 5 deg.) is quite high (14.7 deg./day), which drops up to ~ 13.6 deg./day towards the poles (in both northern and southern hemispheres). Thus, the rotation rate of equatorial region is highest and drops towards the poles that confirm clear differentiality in rotational profile of transition region during Solar Cycle 24. It is also clear from the figure that latitudinal rotational differentiality is least during high solar activity (rotation rate varies from 14.86 to 14.27 deg./day) and increases during the ascending and the descending phases of the solar cycle 24 (rotation rate varies from 14.56 to ~ 13.56 deg./day in 2008 and 14.6 to 13.1 deg./day in 2018). There is also evidence of hemispheric asymmetry in the rotational profile which seems to have temporal variability.

**Table 2:** The comparison of solar rotational coefficients of present work with other published results.

| Tracers/Method | Time period | $C_0 \pm SE$ | $C_2 \pm SE$ | Source | References |
|---|---|---|---|---|---|
| CBPs | 1 - 2 Jan 2011 | 14.62 ± 0.08 | - 2.02 ± 0.33 | SDO/AIA | Sudar et al. (**2014**) |
| XBPs | 1994 - 1997 | 14.39 ± 0.01 | - 1.91 ± 0.10 | Yohkoh/SXT | Hara (**2009**) |
| CBPs | 1992 - 2001 | 17.59 ± 0.39 | - 4.54 ± 1.14 | Yohkoh/SXT | Kariyappa (**2008**) |
| CBPs | 1998 - 1999 | 14.45 ± 0.03 | - 2.22 ± 0.07 | SOHO/EIT | Brajsa, Wohl & Vrsnak (**2004**) |
| XBPs | Jan, Mar and Apr 2007 | 14.19 ± 0.17 | - 4.2 ± 0.77 | Hinode/XRT | Kariyappa (**2008**) |
| Flux Modulation | 1999 - 2001 | 14.8 ± 0.06 | - 2.13 ± 0.14 | NoRH | Chandra, Vats & Iyer (**2009**) |
| Flux Modulation | 1992 - 2001 | 14.50 ± 0.10 | - 0.8 ± 0.60 | Yohkoh/SXT | Chandra, Vats & Iyer (**2010**) |
| Flux Modulation | 2008 - 2018 | 14.7 ± 0.26 | - 1.26 ± 0.5 | STEREO/304 | Present Work |

.

**3.1 ROTATIONAL COEFFICIENTS: A COMPARATIVE STUDY**

The Table 1 contains the rotational coefficients with corresponding standard errors (SE) and root mean square (RMSE) errors (c.f., last column). From Table 1, it is clear that the equatorial rotation rate ($C_0$) doesn't have temporal variability, whereas differential gradient ($C_2$) and asymmetry indicator ($C_1$) show considerable variation from 2008 to 2018.

In the last one and half decade, several researchers have reported the coronal rotation rate based on various techniques using data obtained from ground and space-based solar full disk (SFD) images recorded by various telescopes at different wavelengths. The comparison of average rotational coefficients with their standard errors (SE) is listed in the Table 2. The comparison of equatorial rotation rate ($C_0$) shows that our results are in good agreement with the results reported by the other authors as listed in the Table 2, except Kariyappa (**2008**) using CBPs of Yohkoh/SXT. To summarize, the coefficient $C_0$ in our case is found to be higher by 0.5 % as compared to the $C_0$ reported by Sudar *et. al.* (**2014**), 2.6 % higher than the Hara (**2009**), 1.7 % higher than the Brajsa *et. al.* (**2004**), 1.37 % higher than Chandra, Vats & Iyer (**2010**), and 3.46 % higher as compared to the Kariyappa (**2008**) using XBPs of Hinode/XRT. Further, there is a considerable decrement (about 18 %) in equatorial rotation rate obtained in our study as compared to the value reported by Kariyappa (**2008**) using CBPs of Yohkoh/SXT. Overall comparison concludes that equator of transition region rotates faster than that of other regions of the solar corona as depicted by results in Table 2. The rotational gradient ($C_2$) of the transition region has low values as compared to that of the different regions of solar corona except Chandra, Vats and Iyer, (**2010**).

Thus, our overall conclusion is that the transition region rotates more rigidly as compared to the corona.

As a matter of fact, the asymmetry indicators $C_1$ (at middle latitudes) in Eq. (3) also play an important role on the investigations of asymmetric behavior of solar rotational profile. According to Vats and Chandra (**2011**), the obtained values of these coefficients are $C_1$ = (0.030, 0.162, - 0.022) for the observations from NoRH/17 GHz (Solar Cycle 23), Yohkoh/SXT (Solar Cycle 23) and Yohkoh/SXT (Solar Cycle 22), respectively. In the present study, average asymmetry indicator at middle latitudes is 0.04 for Solar Cycle 24, which is comparable to that of Vats and Chandra (**2011**) for Solar Cycle 22 using NoRH/17 GHz. However, it is much lower as compared to the observations from the Yohkoh/SXT (Solar Cycle 23) (Vats and Chandra (2011). This clearly shows that the asymmetric profile of the solar transition region at the middle latitudes has almost same behavior as that of the corona during the Solar Cycle 22.

**3.2 TREND OF AVERAGE ROTATION RATE**

In order to calculate the average rotation rate of the solar transition region over the full disk, we take the average of all the flux values in the daily images at 30.4 nm. For this purpose, new time series for each year is obtained. This averaged flux contains the signals of all features over the solar disk and as a result, average rotation rate is obtained with high level of accuracy.

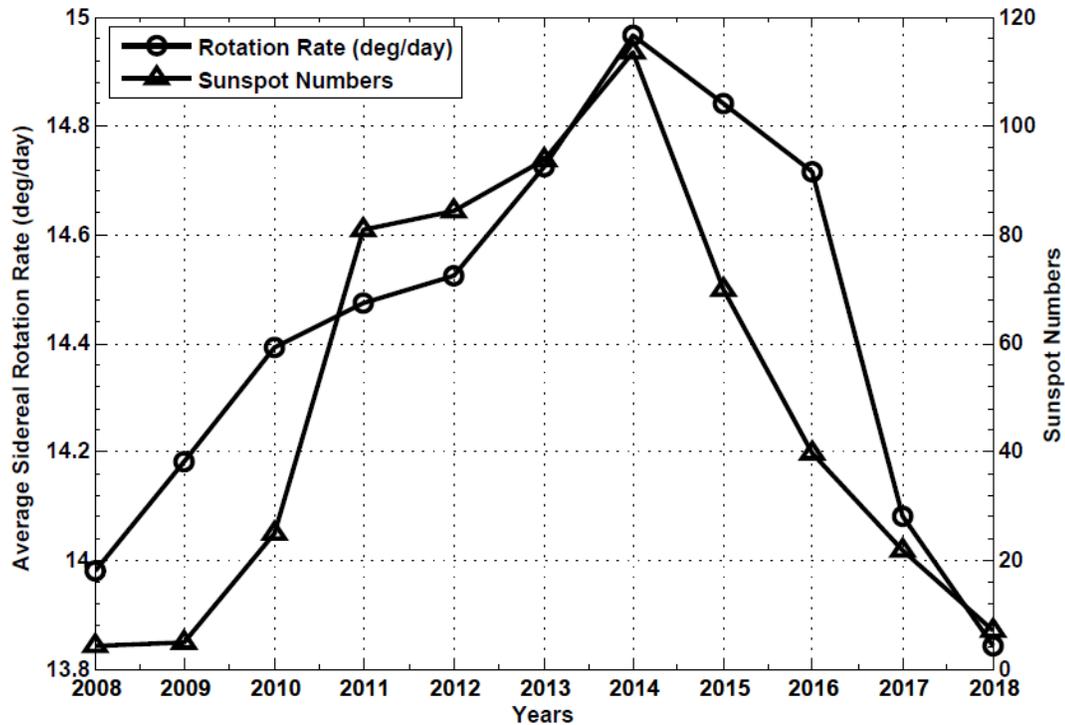

**Figure 5:** Temporal variation of average rotation rate (deg./day) and annual sunspot numbers during the period from 2008 to 2018 (Solar Cycle 24).

In Fig. 5, temporal variation of average rotation rate (deg./day) and annual sunspot numbers corresponding to Solar Cycle 24 from 2008 to 2018 are shown. The obtained average rotation rate of transition region as shown in Fig. 5 is highest in the year 2014 (14.97 deg./day) and it slowly decreases up to 13.98 deg./day and 13.84 deg./day, respectively on both sides of the solar maximum phases. Fig. 5 also shows that the average rotation rate of transition region and sunspot numbers are almost in phase. Furthermore, it appears that average rotation rate of transition region follows the solar activity cycle from 2008 to 2018.

### 3.3 DEPENDENCY OF ROTATION RATE OF TRANSITION REGION ON SOLAR ACTIVITY

To investigate the dependence of the rotation rate of the solar transition region on solar activity, we cross-correlate the average rotation rate (deg./day) of the full disk images of the Sun with annual sunspot numbers at the lag of - 5 to + 5 (in years) for the period from 2008 to 2018 (Fig. **6**).We obtained highest value of the cross-correlation coefficient as 0.86 at zero lag, which clearly shows a strong correlation between aforementioned parameters. To ascertain the statistical significance of our cross-correlation analysis, we use t-test. A very significant (99.9%) result regarding the follow-up of overall rotation rate (deg./day) with respect to solar activity is found.

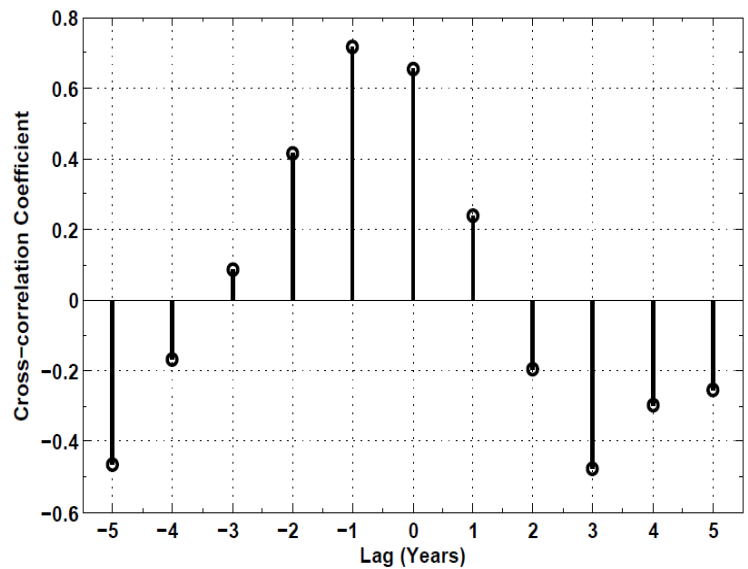

**Figure 6:** Above stem plot shows cross-correlation of average rotation rate (deg./day) of the solar full disk for the transition region (30.4 nm) and annual sunspot numbers as a function of lag (years).

Our results show that average rotation rate of the solar transition region follows the solar activity in a similar way as the relation between rotations and $H_\alpha$ emission activity found in F8-G3 dwarfs (Mekkaden, **1985**).

## 4 CONCLUSIONS

The analysis of *STEREO-A* full-disk observations obtained at 30.4 nm for the solar transition region for the period from 2008 to 2018 depicts the differential nature in rotation rate. The sidereal rotation rate is maximum at the equatorial region and it has decreasing trend on both sides towards the poles. Moreover, the differentiality is least during high solar activity period (rotation rate varies from 14.86 to 14.27 deg./day) whereas increases during the ascending and the descending phases of the solar activity cycle (rotation rate varies from ~ 14.6 to 13.1 deg./day).The average gradient of transition region is seen to be smaller in comparison to the different layers of the solar corona as illustrated in the Table **2** (except Chandra, Vats and Iyer, **2010**). Hence, we can conclude that on an average, the solar transition region rotates less differentially as compared to the different layers of the solar corona. The reason for this rigidness is largely unknown. It could be owing to the higher density of the transition region than that of the corona. However, further investigations to understand the rigidness of solar rotation from interior to the outer atmosphere of the Sun are required. Our results show very strong correlation (0.86) between the average rotation rate of the full disk images of the solar transition region and the solar activity (sunspot numbers).It is also noteworthy to mention that the average rotation rate of transition region follows the solar activity during Solar Cycle 24in the same way as that observed in F8-G3 dwarfs (Mekkaden, **1985**) using $H_\alpha$ emission as stellar activity.

## ACKNOWLEDGEMENTS

The authors wish to acknowledge the use of *STEREO*-30.4 nm observations for the period from 2008-2018. These are acquired from the webpage of *STEREO*, a mission of National Aeronautics and Space Administration (NASA). We also acknowledge the webpage of SILSO-SIDC from which annual sunspots numbers from 2008 to 2018 have been used in this analysis. The research at Udaipur Solar Observatory (USO), Physical Research Laboratory, Udaipur is supported by Department of Space, Government of India. JS and AKM acknowledge the various supports for this research work provided by Department of Physics, CCS University, Meerut. We are highly grateful to the referee for useful comments and suggestions to improve the contents and presentation of the manuscript.

## DATA AVAILABILITY

The data underlying this article will be shared on reasonable request to the corresponding author.